\documentstyle[preprint,revtex]{aps}
\begin{document}
\begin{title}
Collective transport in arrays of quantum dots
\end{title}

\author{A. Alan Middleton and Ned S. Wingreen}
\begin{instit}
NEC Research Institute, 4 Independence Way, Princeton, NJ 08540
\end{instit}
\receipt{\today}

\begin{abstract}

Collective charge transport is studied in one- and
two-dimensional arrays of
small normal-metal dots separated by tunnel barriers.
At temperatures well below the charging energy of a dot,
disorder leads to a threshold for conduction which grows
linearly with the size of the array.
For short-ranged interactions, one of the correlation length
exponents near threshold is found from a
novel argument based on interface growth.
The dynamical exponent
for the current above threshold is also predicted
analytically, and the requirements for its experimental
observation are described.

\end{abstract}

\pacs{
        73.40.Rw, % Metal-insulator-metal structures
        05.60.+w, % Transport processes: theory
        73.40.Gk  % Tunneling in interface structures
        }

%\narrowtext

Systems exemplifying collective transport
in quenched disorder include
sliding charge-density waves (CDW's) \cite{CDW,NF}, fluids in
disordered media \cite{Robbinsetc}, and type-II
superconductors \cite{colltrans}.
For these dynamical systems, there does not yet exist
a classification whereby the long-wavelength behavior
can be predicted from the characteristics of
the microscopic degrees of freedom.
To study this question of universality
experimentally requires systems where
the microscopic degrees of freedom, the
range of interactions, and the nature of the disorder
are well understood.
Here, we propose as a model system an array of
small-capacitance normal-metal dots.

In this letter we examine the low-temperature, nonlinear
charge transport in such an array.
The dots are treated as capacitively coupled conductors with
charges allowed to tunnel between neighboring dots.
In contrast with previous work \cite{bakh},
we explicitly include the
effects of random offset charges on each dot
and investigate the
limit where the number of dots becomes large.
We find that the onset of conduction
occurs at a voltage $V_T$ proportional to the linear array size.
One of the correlation
lengths that diverges near this threshold is found
from a general argument based on interface growth, while
another is found by focusing on ``slow points'' which control
the current.
These correlation lengths determine the
branching of current paths in the array and
hence the current near onset.
In particular, we predict that the current through
linear and square arrays behaves as
 \begin{equation}\label{zeta}
I \sim (V/V_T-1)^{\zeta}
 \end{equation}
with $\zeta=1,5/3$ in dimensions $d=1,2$, respectively.

The array we study is depicted in
 Fig.\ \ref{arraypict}.
For a tunneling resistance $R$ between dots
large compared to the quantum resistance $h/e^2$,
the state of the
array is fully described by the number of electrons in each dot.
The energy is then all electrostatic and is determined by a
matrix of capacitances $C_{ij}$.
We assume a constant capacitance $C$ between neighboring dots and
between the leads and adjacent dots,
and a capacitance $C_g$ between
each dot and the back-gate which underlies the entire array.
The leads and back-gate are taken to have infinite
self-capacitance.
We concentrate on the Coulomb-blockade regime, where
the thermal energy is much smaller than
the charging energies, i.e.,
$k_{B}T \ll e^2/\left[2\,{\max(C,C_g)}\right]$
(for $1\mu\rm m$ dots, this energy is
$\sim 1 {\rm meV}$ \cite{expt}).
We measure distances in units of the dot spacing.

Given the charge $Q_{i}$ on each dot, the electrostatic energy
is \cite{bakh}
 \begin{equation}
E = \sum_{{\rm dots}\ i,j} (Q_{i}+q_{i})C^{-1}_{ij}(Q_{j}+q_{j})
+ V_{L}Q_{L} + V_{R}Q_R{}
+\sum_{{\rm dots}\ i} (V_{iL}+V_{iR}-V_{g})Q_{i},
 \end{equation}
where $Q_{L,R}$ are the charges on the leads,
at voltages $V_{L,R}$,
and $V_g$ is the back-gate voltage.
Disorder is included through the offset
charges $q_{i}$ which represent the effective charge
on each dot due to nearby charged impurities.
Large fluctuations in disorder will be compensated
by an integral number
of mobile charges, so that $0 \le q_{i} < e$.
The voltage on dot $i$ due to the
left(right) lead is given by $V_{iL(R)} = C^{-1}_{iL(R)}V_{L(R)}$.
In general, the elements of the inverse capacitance matrix
fall off exponentially with a screening length $\lambda$
that increases with $C/C_g$
(for $C >> C_g$, $\lambda \approx (C/C_g)^{1/2}$).

At low temperatures, a charge may tunnel between dots
only if such an event lowers the
electrostatic energy of the array.
The kinetic energy gained by the tunneling
electron is assumed to be dissipated
\cite{footcotun}.
The tunneling rate from one configuration
$S=(..., Q_{i} , ... , Q_{j} , ...)$ to another configuration
$S'=(..., Q_{i}-1, ... , Q_{j}+1 , ...)$, where $i$ and $j$ are
neighboring dots or a dot and a neighboring lead is given by
 \begin{equation}\label{rates}
\nu_{S\rightarrow S'}  =
(e^2R)^{-1}\,
\theta\left(E(S)-E(S')\right)\left[ E(S) - E(S') \right].
 \end{equation}
This rate grows linearly with energy gain since
the number of electrons available to tunnel is proportional
to the relative shift of the Fermi surfaces in the dots.
For arrays of a few junctions, numerical results on
this model \cite{bakh}
compare well with experiment \cite{expt}.

For large arrays, we find a second-order
transition, with associated critical phenomena,
which separates a static, non-conducting state from a dynamic,
conducting state.
The control parameter is the voltage
difference between the leads. At low voltage
differences, the array always relaxes to a static configuration,
while at high voltage differences,
charges traverse the array from one lead to the other.

An important
question to ask is whether the conduction transition
is hysteretic for a given realization of disorder.
In one-dimensional systems, the current is a unique function of the
applied voltages, regardless
of the magnitude of $\lambda$ \cite{uniq}.
In two-dimensional arrays at zero temperature, the current
{\em can} depend on the history of the applied voltages.
To within our numerical accuracy, however,
the current is history independent in typical samples.
It can furthermore be shown that the current
is entirely independent of history
in the limit of short screening lengths,
$C/C_g \rightarrow 0$, to which we shall
devote most of our attention.

We have numerically determined the dependence of the
threshold voltage for conduction, $V_{T}(N) = V_{L}-V_{R}$,
on the ratio $C/C_g$ and on the linear system size $N$
(for fixed gate voltage $V_{g}=0$).
We find that the threshold voltage is proportional to $N$,
 \begin{equation}
\lim_{N\rightarrow\infty} \overline{V_T}(N)C_{g}/Ne
= \alpha(C/C_g),
 \end{equation}
where the overbar represents an average over disorder.

The function $\alpha(C/C_g)$ for one-dimensional
arrays is plotted in Fig \ref{arraypict}.
In the limit $C/C_g \rightarrow 0$,
the voltage on a dot is just $(Q_{i}+q_i)/C_{g}$, as the
capacitive coupling between dots is negligible.
The schematic in  Fig.\ \ref{slopes}(a) shows that in order to carry
a current in this limit, the voltage difference across the
array must be large enough to overcome $\approx N/2$ upward steps
in the random potential.
This observation gives
$\alpha(C/C_g\rightarrow 0) = 1/2$.
In the limit of large $C/C_g$ (large $\lambda$),
$\overline{V_{T}}$ can be estimated
by balancing the ``force'' on the charges due to
a charge density gradient against the random potential
gradient \cite{SCbean}.
It is necessary to
recognize that there is a stability limit for the dot-to-dot
potential difference: at higher potential differences, charges
will tunnel, reducing the potential
across the tunneling barrier.
Estimating the magnitude of the pinning forces to be given by
this stability limit, we find \cite{tobepubed}
a maximum static density
gradient of $\sim e/\lambda^{2}$, that is, a
density change of $O(1)$ charge per screening length is allowed
in regions separated by $\lambda$.
This gives $\alpha \sim (C/C_g)^{-1}$ at large $\lambda$.
Numerically, we find in $d=1$ that
$(C/C_g) \alpha \rightarrow 0.10(1)$,
as $C/C_g \rightarrow \infty$.

We now discuss the approach to the conduction
threshold in two dimensions,
in order to elucidate the critical behavior of the
correlation lengths and current.
We concentrate on the limit
$C/C_g \rightarrow 0$, that is, $\lambda$ small, in order
to maximize the number of effective degrees of freedom
and to eliminate hysteresis.
%The random potential on each dot is $q_{i}/C_{g}$.
Taking $V_{R}$ to be fixed and raising $V_{L}$, charge moves from
the left lead onto the array.
The condition for a charge to overcome the Coulomb barrier
and tunnel from site $i$ to neighboring site $j$ is
 \begin{equation}
\label{hubbcondition}
V_{i} > V_{j} + e/C_g.
 \end{equation}
At a given $V_{L}$, this advance of charge is halted when
$V_{i} \le V_j+e/C_g$ everywhere.
Though the tunneling is stochastic,
the static configuration at any $V_{L} < V_{T}$ is
entirely determined by the disorder realization.
The distance to which charge penetrates therefore
defines a unique interface
(given $Q_i=0$ initially).
When $V_{L}$ is raised by $e/C_{g}$, charge is added to each
point on the interface and {\em therefore the interface must
advance by at least one lattice spacing}.
In addition, the interface may advance
further at some points if the local disorder is favorable.
The motion of this interface is depicted in
 Fig.\ \ref{slopes}(b);
conduction occurs when the interface reaches the right lead.
Numerical calculations of this threshold give
$\alpha(C/C_g\rightarrow 0)= 0.338(1)$ in $d=2$.

The dynamics of the Coulomb-blockade
condition  Eq.\ (\ref{hubbcondition})
make the interface ``motion'' with increasing $V_L$
similar to the stochastic growth of interfaces
in models without quenched spatial disorder,
such as the one due to Eden \cite{Eden}.
%(commented out)
%Though the interface is moving through a quenched impurity
%potential, the required advance at each step implies that the
%effects of the disorder have only short-range correlations
%in $V_L$.
The results on the Kardar-Parisi-Zhang (KPZ) equation for
a $(d-1)$-dimensional interface \cite{KPZ} subject
to short-range correlated noise are therefore useful in
understanding the behavior of a $d$-dimensional array
of dots.
This is to be contrasted with the usual motion of interfaces at
small velocities through random media, where the interface can
be pinned for some time at one point, resulting in long time
correlations
and exponents distinct from those of the KPZ equation \cite{corrs}.

In the case $d=2$, the results for KPZ interfaces imply
that the width of the interface must scale as
$V_{L}^{1/3}$ \cite{KPZ}.
Furthermore, the fluctuations in the position of maximum advance
of the interface behave as $\sim V_L^{1/3}(\ln V_L)^{1/2}$.
The rms fluctuations $\Delta V_T$ in $V_T$
and the mean threshold voltages $\overline{V_T}$
therefore behave as
 \begin{eqnarray}
\label{scaletwo}
\Delta V_T/\overline{V_T}  \sim N^{-2/3}(\ln N)^{1/2},\\
\label{scaleone}
\overline{V_T}(N)C_g/eN-\alpha \sim N^{-2/3}(\ln N)^{1/2},
 \end{eqnarray}
for an $N\times N$ array \cite{square}
(for $d=1$, $\Delta V_T/V_T \sim N^{-1/2}$).
The fluctuations in the threshold voltage as a function of
size may be used to define a finite-size scaling exponent
$\nu_T$ via $N \sim (\Delta V_T/ \overline{V_T})^{-\nu_T}$.
This length, besides giving
the fluctuations in $V_T$, determines the finite
size crossover in quantities such as the polarization of
the array \cite{NF}.
{}From  Eq.\ (\ref{scaletwo}), we find $\nu_T = 3/2$.
Numerical simulations on systems up to size $N = 2560$
are fit very well by  Eq.\ (\ref{scaletwo}),
as shown in the inset in  Fig.\ \ref{slopes}(a),
and by  Eq.\ (\ref{scaleone}).

The current in the one-dimensional model with only on-site
interactions can be understood in detail.
For voltages much greater than threshold,
$v \equiv (V_L-V_T)/V_T \gg 1$, the charge gradient
across each junction is much greater than one.
By  Eq.\ (\ref{rates}), the current is then approximately
 \begin{equation}\label{onedcurr}
I \approx (e/2RC_g)v.
 \end{equation}
In contrast, near threshold, the discreteness
of the charges and the disorder become important.
The excess charge gradient above the threshold
configuration is composed of
steps that occur at well separated ``slow points'',
located where the potential drop between dots in
the threshold configuration is small compared to $e/C_g$.
It can be shown \cite{tobepubed} that the current is given by
the fastest ``slow point''; the tunneling rate across this
point is, on average, $(V-\overline{V_T})/eRN$.
Interestingly, near threshold this also gives
 Eq.\ (\ref{onedcurr}).
We therefore find $\zeta=1$ for  Eq.\ (\ref{zeta}) in $d=1$.
As shown in  Fig.\ \ref{twodimflow}(a),  Eq.\ (\ref{onedcurr})
is consistent with our numerical results near and far
from threshold.

The pattern of current flow in a typical
two-dimensional array is shown in  Fig.\ \ref{twodimflow}(b).
At voltages just above threshold, $V_L-V_T << e/C_g$, the current is
in general carried on a single path, with little or no branching.
This path is exactly
the one with the minimal number of upward steps
in the potential between the two leads.
Previous work \cite{KPZ} shows that such paths have
transverse fluctuations
$\sim n^{2/3}$, where $n$ is the distance from the
left lead;
this is consistent with our numerical results.
Increasing the voltage to a few times $e/C_g$ above threshold
opens multiple channels
which branch and reconnect, as shown in  Fig.\ \ref{twodimflow}(b).
Note that at voltages $V_L$ exceeding threshold by
$O(eN^{1/3}(\ln N)^{1/2}/C_g)$,
current can in principle flow anywhere in the array, as
the charge invasion interface contacts the right lead at each point.
However, near threshold, all but a small fraction of
the current is confined to a few major current-carrying paths.
The selection of these
paths out of all possible paths results from a characteristic
length for branching, as we now discuss.

Any current-carrying channel between the two leads
must have $(V_L-V_T)C_g$ excess steps in the charge
density relative to
the threshold configuration.
This gives a correlation length
$\xi_\parallel = eN/(V_L-V_T)C_g$ which separates the steps in the
excess charge, as in $d=1$.
This length determines the separation between branch
points along the channels \cite{tobepubed}.
The channels therefore
wander transversely a distance $\xi_\parallel^{2/3}$
between branch points,
giving a channel separation of $\xi_\perp \sim v^{-2/3}$.
The current through each channel
behaves as $\sim (e/2RC_g)v$, since each segment between branch
points is one-dimensional.
The current through the array is then given by
 \begin{equation}
I \sim (e/2RC_g) v N/\xi_\perp \sim (e/2RC_g)N v^{5/3},
 \end{equation}
resulting in $\zeta(d=2) = 5/3$.

Our numerical results for the transport in
two-dimensional systems are
shown in  Fig.\ \ref{twodimflow}(b).
The current-voltage relationship is approximately
fit by $I/N \sim v^{2.0}$,
over the range $10^{-2} < v < 10^{-1}$, but the slope on
a log-log plot does not converge in the range we have
studied numerically.
To observe the true exponent requires arrays larger
than $400^2$, either numerical or experimental.

In conclusion,
we have determined the threshold for conduction in
arrays of small
normal-metal dots with disorder.
By examining correlation lengths that describe the separation
of parallel current paths and the distance between
dynamically important ``slow points''
we have determined the transport behavior near the threshold.
The critical exponents for the current and correlation
lengths
which we have derived using the KPZ interface model
are distinct from those found for elastic media \cite{CDW} and
fluid flow \cite{Robbinsetc,rivers}.
These differences are clearly related to the
novel features of this system, namely the
discreteness of the carriers and (quantum) stochastic flow,
which result in (a) an always advancing charge interface below
threshold and (b) the non-local selection of
current paths above threshold.

We wish to thank Paul McEuen for encouraging our interest
in arrays, and Chao Tang for many valuable discussions.

\figure{\label{arraypict}
The threshold voltage per dot, $\overline{V_{T}}/N$,
in units of $e/C_{g}$, for conduction through
a one-dimensional array of normal-metal dots
as a function of $C/C_{g}$.
The dashed lines show analytical predictions.
A two-dimensional array of dots is shown in the inset.
The indicated capacitance between dots
is $C$, while the capacitance between each dot and the
back-gate (the dashed rectangle) is $C_{g}$.
At $T=0$, charges may only tunnel between neighboring dots
if this lowers the total electrostatic energy.
The voltages applied to the left lead, the right lead, and the
gate are indicated as $V_{L}$, $V_{R}$, and $V_{G}$,
respectively.
}

\figure{\label{slopes}
(a) Schematic of dot voltages for a
one-dimensional array below the
threshold for conduction, in the limit of short
screening length $\lambda$.
Each square indicates an increase in
the on-site voltage by $e/C_{g}$
due to an added charge; the relative offset in the voltages
is caused by quenched disorder.
If the left-lead voltage $V_{L}$ is further raised by $e/C_{g}$,
charges will tunnel onto the array until stopped by the
next upward step, as indicated
by the dashed squares.
The inset shows the calculated sample-to-sample fluctuation
of threshold voltages, $V_T$, in one and two dimensions
as a function of linear system size, $N$, with fits described
in the text ( Eq.\ (\ref{scaletwo})).
(b) Contours of constant charge occupation in a $160^2$
array at the threshold $V_T$ (contour spacing is 5 charges).
Successive contours coincide with the distance to which charge
flows for various voltages below threshold, $V_{L} < V_{T}$.
Conduction occurs at the voltage where the charge first
reaches the right lead.
}

\figure{\label{twodimflow}
(a) Plot of current-voltage relationship near threshold
for one- and two-dimensional arrays of various sizes.
The numbers in parentheses give the number of disorder
realizations.
For one-dimensional arrays, the current both near
and far from threshold is well fit
by  Eq.\ (\ref{onedcurr}).
The data for the $d=2$ arrays
are approximately fit by $I \sim (V-V_T)^{2.0}$
at the lowest currents shown, though the local slope
on the log-log plot has not converged.
(b) Current paths in a two-dimensional array (of size $160^2$)
at two voltages near threshold.
Very near threshold, the current flows in a single narrow
channel (dark line).
Multiple, branching channels are shown for
a voltage $3e/C_g$ above threshold (light lines).
}


\begin{references}

\bibitem{CDW}
For reviews, see {\it Charge Density Waves in Solids},
edited by L.\ P.~Gorkov and G.~Gr\"uner (Elsevier, Amsterdam,
1989).

\bibitem{NF}
O.~Narayan and D.~S.~Fisher,  Phys.\ Rev.\ Lett. {\bf 68}, 3615 (1992)
and  Phys.\ Rev.\ B {\bf 46}, 11520 (1992); A.~A.\ Middleton and D.~S.\
Fisher,
 Phys.\ Rev.\ B {\bf 47}, 3530 (1992).

\bibitem{Robbinsetc}
N.~Martys, M.~Cieplak, and M.~O.\ Robbins,  Phys.\ Rev.\ Lett. {\bf 66}, 1058
(1991).

\bibitem{colltrans}
D.~S.\ Fisher, in {\it Nonlinearity in Condensed Matter}, edited by
A.~R.\ Bishop, et al. (Springer-Verlag, New York, 1987).

\bibitem{bakh}
U.~Geigenm\"uller and G.~Sch\"on, Europhys.\ Lett.\ {\bf 10},
765 (1989); N.~S.\ Bakhvalov, et al., Sov.~Phys.~JETP {\bf 68},
581 (1989).

\bibitem{expt}
P.~Delsing, in {\em Single Charge Tunneling}, H.~Grabert
and M.~H.\ Devoret, eds. (Plenum, New York, 1992).

\bibitem{footcotun}
For small-capacitance and high-resistance junctions,
the effects of small finite-temperatures and of co-tunneling,
in which electrons tunnel coherently through two barriers,
are minimal [D.~V.\ Averin and Yu.~V.\ Nazarov,
{\it ibid} \cite{expt}].

\bibitem{uniq}
A ``no-passing'' rule for the integrated current through
the junctions can be used to show that, at fixed voltage,
each initial state
has the same average velocity at long times;
see A.~A.\ Middleton,  Phys.\ Rev.\ Lett. {\bf 68}, 670 (1992).

\bibitem{SCbean}
C.~P.\ Bean,  Phys.\ Rev.\ Lett. {\bf 8}, 250 (1962).

\bibitem{tobepubed}
A.~A.\ Middleton and N.~S.\ Wingreen, in preparation.

\bibitem{Eden}
M.~Eden, in {\it Proceedings of the Fourth Berkeley Symposium
on Mathematical Statistics and Probability}, edited by F.~Neyman
(University of California, Berkeley, 1961), Vol.~IV.

\bibitem{KPZ}
M.~Kardar, G.~Parisi, and Y.-C.\ Zhang,  Phys.\ Rev.\ Lett. {\bf 56}, 889
(1986);
D.~A.\ Huse and C.~L.\ Henley,  Phys.\ Rev.\ Lett. {\bf 54}, 2708 (1985);
S.~Roux, A.~Hansen, and E.~L.\ Hinrichsen, J.~Phys.~A {\bf 24},
L295 (1991); L.-H.\ Tang, B.~M.\ Forrest, and D.~E.\ Wolf,
 Phys.\ Rev.\ B {\bf 45}, 7162 (1992).

\bibitem{corrs}
K.~Sneppen,  Phys.\ Rev.\ Lett. {\bf 69}, 3539 (1992), also see
Ref.\ \cite{Robbinsetc}
and references therein.

\bibitem{square}
We cite results here for a square $N\times N$ array. For an
anisotropic array, with a linear distance between the leads
of $M$ and width $N$, the fluctuations
in the threshold voltage scale as $N^{1/2}$,
for $MN^{-2/3} \ll 1$.
This method can also be applied to CDW models:
numerically, the interface width is constant, reflecting
elastic behavior,
but the average velocity has $t^{-d/2}$
fluctuations, leading to the known value $\nu_T = 2/d$
(Ref.\ \cite{NF}).

\bibitem{rivers}
H.~Takayasu,  Phys.\ Rev.\ Lett. {\bf 63}, 2563 (1990);
O.~Narayan and D.~S.\ Fisher, to be published.

%\bibitem{footnote2}
%The behavior at fixed $V$ is statistically identical to that
%at fixed field $V/N$, though the response to a varying
%potential is quite distinct from the response to a varying
%electric field.
%This relationship holds for both the quantum dot arrays and
%CDW's.

\end{references}
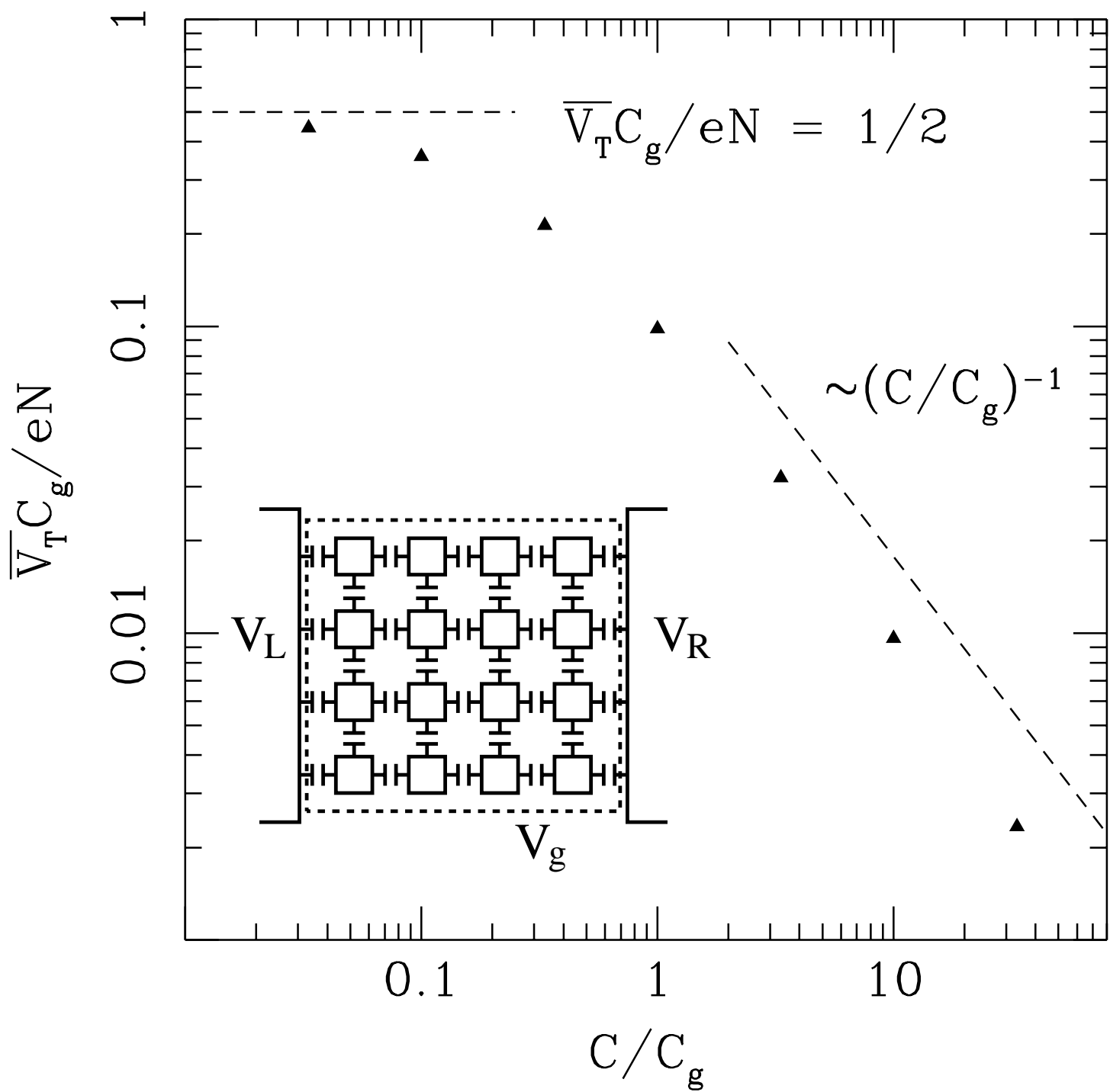
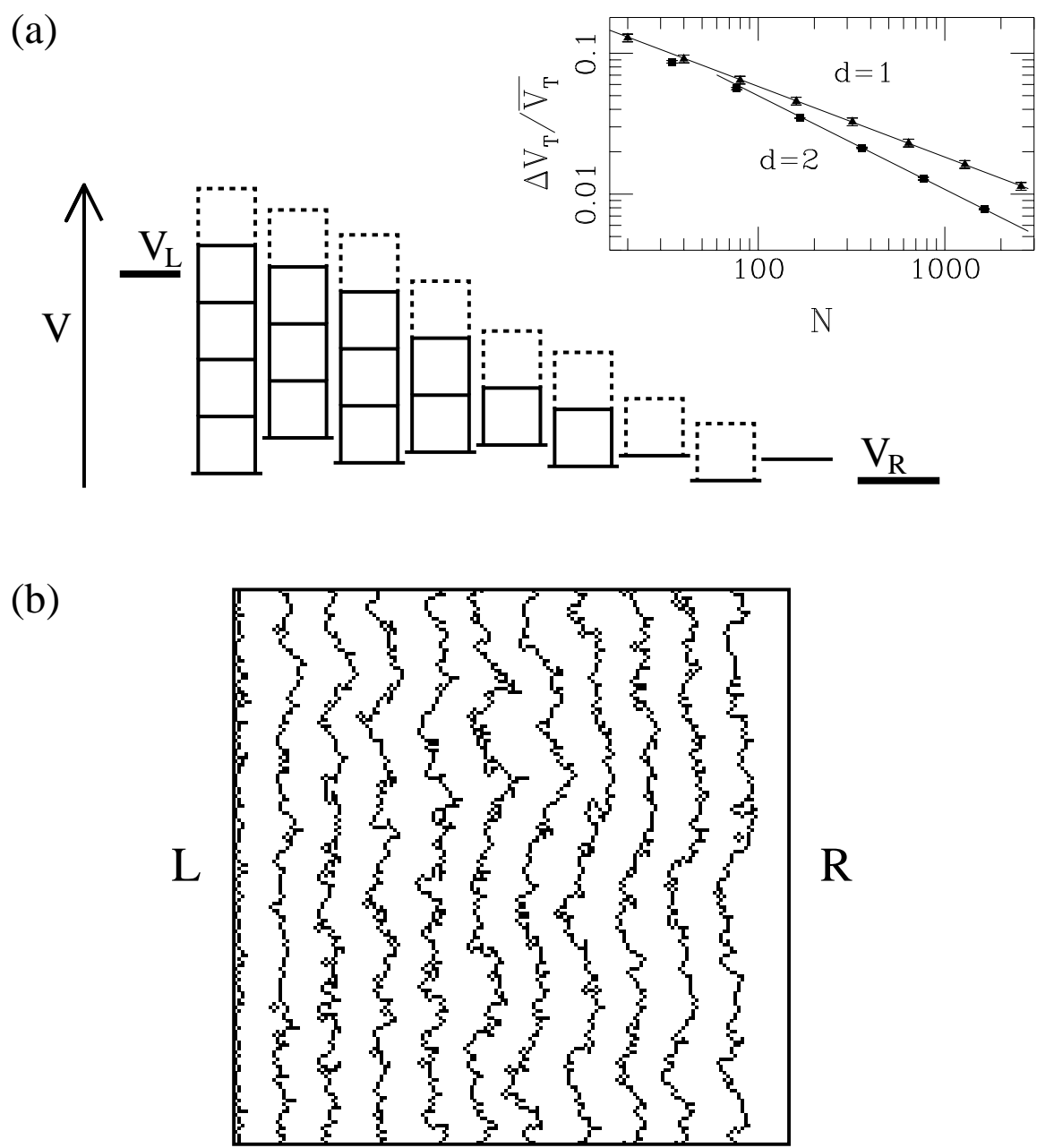
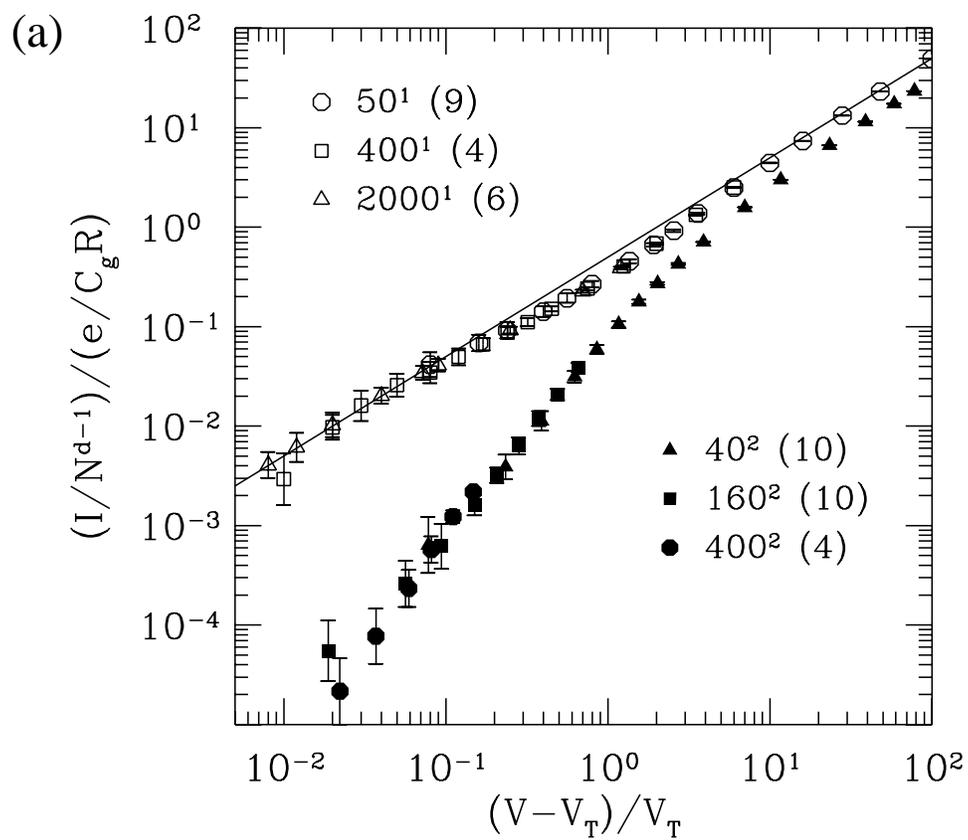
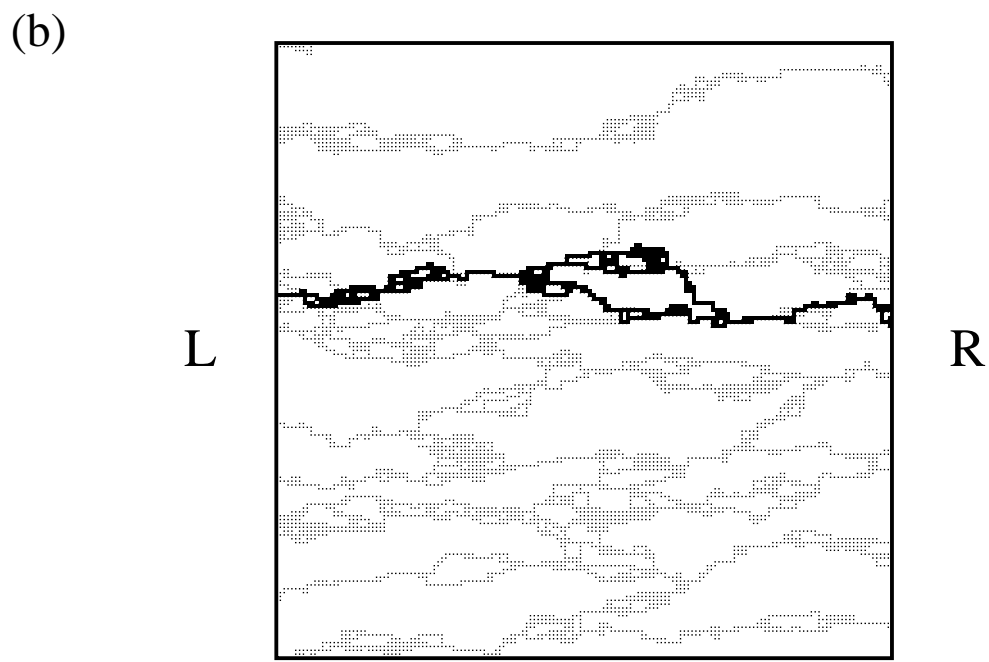
\end{document}